\documentclass[final,twocolumn,5p]{elsarticle}

\usepackage[%
        hyperindex,breaklinks,
	pdfstartview=FitH,
  pdfpagemode=UseNone
	]{hyperref}

\usepackage[
   centertags, 
   sumlimits,  
   intlimits,  
   namelimits, 
]{amsmath} %
\usepackage{amssymb}
\usepackage{nicefrac}
\usepackage{todonotes}

\usepackage[all]{hypcap} 

\usepackage[english]{babel} 
\usepackage{csquotes}

\usepackage[%
]{graphicx}
\usepackage[caption=false]{subfig}

\usepackage[range-phrase=-,per=fraction,fraction=nice,expproduct=cdot,alsoload=hep]{siunitx} 
\newunit{\nuclearNumber}{A}
\newunit{\AGeV}{\nuclearNumber\giga\electronvolt}
\newunit{\ATeV}{\nuclearNumber\tera\electronvolt}
\newunit{\fmc}{\femto\metre\per\clight}
\newunit{\fm}{\femto\metre}
\newunit{\GeVc2}{\giga\electronvolt\per\clight\squared}
\newunit{\GeVc}{\giga\electronvolt\per\clight}
\newunit{\GeVpercubicfm}{\giga\electronvolt\per\cubic\fm}

\biboptions{numbers,square,sort&compress}

\newcommand{\vect}[1]{\boldsymbol{\mathbf{#1}}}

\def\Bamps{\textsc{Bamps}}
\def\Pythia{\textsc{Pythia}}
\def\pp{p\,+\,p}
\def\PbPb{Pb\,+\,Pb}
\def\Raa{$R_{AA}$}

\begin{document}
\title{The different energy loss mechanisms of inclusive and b-tagged reconstructed jets within ultra-relativistic heavy-ion collisions}

\author[ffm]{F.~Senzel\corref{cor1}}
\ead{senzel@th.physik.uni-frankfurt.de}

\author[ffm]{J.~Uphoff}

\author[tsinghua]{Z.~Xu}

\author[ffm]{C.~Greiner\corref{cor2}}

\cortext[cor1]{Corresponding author}
\cortext[cor2]{Principal corresponding author}

\address[ffm]{Institut f\"ur Theoretische Physik, Goethe-Universit\"at Frankfurt, Max-von-Laue-Str.\ 1, 
D-60438 Frankfurt am Main, Germany}

\address[tsinghua]{Department of Physics, Tsinghua University and Collaborative Innovation Center of Quantum Matter, Beijing 100084, China}

\date{\today}

\begin{abstract}
The phenomenon of jet quenching provides essential information about the properties of hot and dense matter created in ultra-relativistic heavy-ion collisions. Recent results from experiments at the Large Hadron Collider (LHC) show evidence for an unexpectedly similar suppression of both light and heavy flavor jets. Furthermore, the role of radiative energy loss of heavy quarks is still under active discussion within the theoretical community. By employing the parton cascade \emph{Boltzmann Approach to Multi-Parton Scatterings} (BAMPS), which numerically solves the 3+1D Boltzmann equation both for light and heavy flavor partons, we calculate the nuclear modification factor of inclusive and b-tagged reconstructed jets in 0-10\% central $\sqrt{s}_{\rm LHC} = \SI{2.76}{\ATeV}$ \PbPb{} collisions. Based on perturbative QCD cross sections we find a suppression of both light and heavy flavor jets. While the inclusive jets are slightly too strong suppressed within \Bamps{} in comparison with data, both elastic+radiative and only elastic interactions lead to a realistic b-tagged jet suppression. To further investigate light and heavy flavor energy loss we predict the $R$ dependence of inclusive and b-tagged jet suppression. Furthermore, we propose the medium modification of b-tagged jet shapes as an observable for discriminating between different heavy quark energy loss scenarios.
\end{abstract}


\maketitle

\pagebreak

As proposed by Bjorken in the early 1980s \cite{Bjorken1982} quenched jets---particles with high transverse momentum $p_t$---represent an excellent probe for investigating the hot and dense matter created in ultra-relativistic heavy-ion collisions called the quark-gluon plasma (QGP) \cite{Braun-Munzinger2007,Jacak2012}. Created under good theoretical control by perturbative quantum-chromodynamics (pQCD) in hard nucleon-nucleon collisions, jets lose energy and momentum while traversing the QGP. Due to their early creation they witness a large part of the QGP evolution and thereby provide essential information about the underlying medium. 

Before the era of the Large Hadron Collider (LHC) at CERN, the most prominent observable for studying jet quenching was the suppression of inclusive hadron spectra quantified by the nuclear modification factor \Raa{},
\begin{align}
\label{raa_def}
  R_{AA}=\frac{{\rm d}^{2}N_{\text{AA}}/{\rm d}p_{T}{\rm d}y}{N_{\rm bin} \, {\rm d}^{2}N_{\text{pp}}/{\rm d}p_{T}{\rm d}y} \, .
\end{align}
Results from both $\sqrt{s}_{\rm RHIC} = \SI{200}{\AGeV}$ Au\,+\,Au collisions at the Relativistic Heavy-Ion Collider (RHIC) \cite{PHENIX2002,STAR2002} and $\sqrt{s}_{\rm LHC} = \SI{2.76}{\ATeV}$ \PbPb{} collisions at the LHC \cite{ALICECollaboration2010a,CMS2012} show a strong suppression of different hadron species at high $p_t$ confirming the picture of quenched jets in heavy-ion collisions.

Due to the increased production cross section of high $p_t$ partons at LHC, complementary studies of jet quenching by reconstructed jets become feasible. Originally developed in elementary collisions like e.g.\ $\text{e}^+\,+\,\text{e}^-$ and \pp{} collisions \cite{Salam2010}, reconstructed jets can provide additional information about the angular dependence of jet energy loss due to their sensitivity to both the leading parton and its surrounding parton shower. Results about the medium modification of reconstructed jets at LHC show both a momentum asymmetry between the leading and subleading jet \cite{ATLASCollaboration2010,CMSCollaboration2011b} together with strong suppression of reconstructed jets measured by the nuclear modification factor \Raa{} of reconstructed jets \cite{ATLASCollaboration2014,ALICECollaboration2015b}. These findings triggered an enormous effort on the theoretical side for describing medium modification of jets \cite{Casalderrey-Solana2011c,Zapp2013,Renk2012,Senzel2015a,Apolinario2013c,VanLeeuwen2015}.

Another promising way for studying jet quenching and especially its mass dependence are energetic charm and bottom quarks. Their larger mass and thereby early production time together with their conserved flavor during hadronization makes heavy quarks a clean tool for examining the QGP. However, the actual energy loss mechanism of heavy quarks is actively debated within theoretical models based on pQCD:  While in Refs.~\cite{Gossiaux2009a, Uphoff2011,Uphoff2012,Young2012a, Meistrenko2013a,Alberico2013a,Uphoff2014} the heavy quarks only scatter elastically and thereby neglect radiative processes, the approaches of Refs.~\cite{Sharma2009,Abir2012a,Buzzatti2012a,Cao2013,Djordjevic2014,Nahrgang2014,Renk2014c,Uphoff2014} also incorporate the possibility of radiative heavy quark processes.

In this letter we combine recent efforts in understanding the medium modification of both heavy quarks \cite{Uphoff2014} and reconstructed jets \cite{Senzel2015a} within the partonic transport approach \Bamps{} (Boltzmann Approach to Multi-Parton Scatterings). To this end, we study the \Raa{} of inclusive and b-tagged reconstructed jets in 0-10\% \PbPb{} collisions at $\sqrt{s}_{\rm LHC} = \SI{2.76}{\ATeV}$. B-tagged jets are defined as jets close in angle $\Delta R = \sqrt{\Delta\phi^2+\Delta y^2}$ to a measured B meson corresponding to bottom quarks and their surrounding shower traversing the medium. As predicted by analytical calculations based on pQCD energy loss \cite{Huang2013}, first results by the CMS collaboration \cite{CMSCollaboration2013} showed evidence for a suppression of b-tagged jets that is unexpectedly similar to the inclusive jet \Raa{}. These findings could be additionally reproduced in other pQCD energy loss models \cite{Djordjevic2016}. While the studies in Ref.~\cite{Huang2013} attribute the similarity of inclusive and b-tagged jets at high $p_t$ to the dominance of the  Landau-Pomeranchuk-Migdal effect, the model of Ref.~\cite{Djordjevic2016} even further predicts a larger suppression of single B mesons in comparison with charged hadrons caused mainly by the different fragmentation into hadrons. To extend these calculations and thereby further discriminate between different heavy quark energy loss mechanisms, we predict within this paper the medium modification of jet shapes \cite{CMSCollaboration2013b}, defined as
\begin{align}
  \rho\left(r\right) = \frac{1}{\delta r} \left< \frac{\sum\limits_{\text{partons}~\in~[r_{\mathrm{a}}, ~r_{\mathrm{b}})}{p_\text{T}^\text{parton}}}{p_\text{T}^\text{jet}} \right>_{\rm jets} \,
  \label{eq:jetshapes}
\end{align}
for reconstructed b-tagged jets in LHC collisions and compare them with results for the inclusive jet shapes by CMS \cite{CMSCollaboration2013b}. After normalizing the jet shapes between $r\in[0,R=\num{0.3}]$ to unity, $\rho(r)$ represents the fraction of jet momentum in an angle between $r-\delta r/2$ at $r$  and $r+\delta r/2$ to each jet axis.

The partonic transport approach \Bamps{} describes the QGP created in ultra-relativistic heavy-ion collisions by numerically solving the 3+1\,D Boltzmann equation,
\begin{align}
p^{\mu} \partial_{\mu} f(\vect{x}, t) = \mathcal{C}_{22} + \mathcal{C}_{2 \leftrightarrow 3} \, ,
\end{align}
for gluons as well as light and heavy quarks by employing a stochastic test-particle Ansatz \cite{Xu2005}. While both the gluons and light quarks (flavor u,d,s) are treated as massless particles, the masses of heavy quarks (flavor c,b) are set to $M_c=\SI{1.3}{\GeV}$ and $M_b=\SI{4.6}{\GeV}$. By evaluating the running of the QCD coupling on a microscopic level, \Bamps{} considers both elastic and radiative Bremsstrahlung processes: elastic matrix elements that are derived from leading-order pQCD and inelastic matrix elements calculated in an improved Gunion-Bertsch approximation \cite{Fochler2013} that was recently extended also to massive particles \cite{Uphoff2014}. This procedure allows an equal treatment of massless and massive partons while any potential mass effect results directly from the underlying pQCD matrix element. Thereby, as shown in Ref.~\cite{Abir2012,Uphoff2014}, the dead cone effect corresponding to a suppression of collinear gluon emissions by a heavy quark is implicitly considered for the full phase space generalizing the seminal work of Ref.~\cite{Dokshitzer2001}. 

Furthermore, the Landau-Pomeranchuk-Migdal effect \cite{Migdal1956} corresponding to a suppression of coherent gluon emissions is effectively treated via a theta function $\theta\left(\tau-X_{\rm LPM}\lambda\right)$ in the inelastic matrix elements, where $\tau$ is the formation time of the emitted gluon and $\lambda$ is the mean free path of the emitting parton. The free parameter $X_{\rm LPM}$ interpolating between the two extreme cases of no LPM suppression ($X_{\rm LPM} = 0$) and the Bethe-Heitler limit of partonic energy loss ($X_{\rm LPM} = 1$) is fixed within this letter to $X_{\rm LPM}=\num{0.3}$ by comparing with the neutral pion \Raa{} data from RHIC \cite{Uphoff2015a,Uphoff2014}. We showed in Ref.~\cite{Uphoff2015a} that this parameter choice results in a realistic suppression of hadron spectra at both RHIC and LHC while the same interactions also build up a sizable elliptic flow $v_2$ in the partonic phase. For more details about the \Bamps{} framework and recent results we refer to Refs.~\cite{Xu2005,Xu2007,Uphoff2015a,Uphoff2014}.

Reconstructed jets are sensitive to the medium modification of both the leading parton and its surrounding shower. Following previous studies about the momentum imbalance $A_J$ of reconstructed jets \cite{Senzel2015a}, we employ the event generator \Pythia{} for the initial distribution of parton showers. For the b-tagged jet results we select only events that contain at least one b-quark. Possible hard processes within \Pythia{} leading to b-quarks in the final state are heavy-flavor creation, e.g. $q\,\bar{q} \rightarrow Q\,\bar{Q}$, heavy flavor excitation, e.g. $g\,Q \rightarrow g\,Q$, and gluon splittings, e.g. $g\,q \rightarrow g\,q \rightarrow q\,\bar{q}\,q$. Studies \cite{Huang2013,CMSCollaboration2011c,CMSCollaboration2013} have shown that especially the contribution of gluon splittings to the production of heavy quarks at higher momentum is significant. All partons from a given shower pair are created at the same point that is spatially distributed by a Glauber modeling with a Woods-Saxon density profile. After their formation time $\tau_f = \frac{2 \omega}{k_t^2}$ the shower partons are subsequently evolved within expanding \Bamps{} simulations of $\sqrt{s}_{\rm LHC} = \SI{2.76}{\ATeV}$ Pb\,+\,Pb collisions with impact parameter $b = \SI{3.6}{\fm}$ corresponding to 0-10\% centrality.  

Based on the medium-modified parton showers, jets are reconstructed by the anti $k_t$ algorithm as provided in the package FastJet \cite{Cacciari2012} while employing different jet radii $R$. Requiring the same trigger and reconstruction conditions as the experiments, the infra-red safety of the anti $k_t$ algorithm allows a reliable comparison of our parton-level calculations to the measured hadron-level jet data. A b-tagged jet is defined as a reconstructed jet whose axis is closer than $\Delta R < R$ to a (anti-)bottom quark.

The shower partons---gluons and light quarks---are allowed to scatter both elastically and via Bremsstrahlung processes. However, to further investigate the underlying heavy quark energy loss mechanism we define different scenarios for the b-quark interactions: Consistent with the light partons, in scenario "el.+rad." b-quarks may scatter both elastically and radiatively. In contrast, in scenario "el." the b-quarks may scatter only via elastic processes. This discrimination allows studying the consequences of medium-induced gluon radiation off the b-quark for the subsequently reconstructed jets. Furthermore, in order to estimate any mass effect within \Bamps{} we initialize in scenario "el.+rad.,${\rm m}_{\rm HQ} = 0$" a b-quark shower within \Pythia{} but consider the bottom quark as massless during the subsequent \Bamps{} evolution. Finally, in scenario "scaled el." we follow our previous studies ~\cite{Uphoff2014} by employing only elastic heavy quark interactions with a modified Debye screening ($\kappa=0.2$) together with a scaling factor $K=3.5$. As discussed in Ref.~\cite{Uphoff2014} these scaled elastic interactions describe both the heavy flavor \Raa{} and $v_2$, while the interactions of scenario "el.+rad." underestimate the flow of heavy quarks.

\begin{figure}[tb]
\includegraphics[width=\columnwidth]{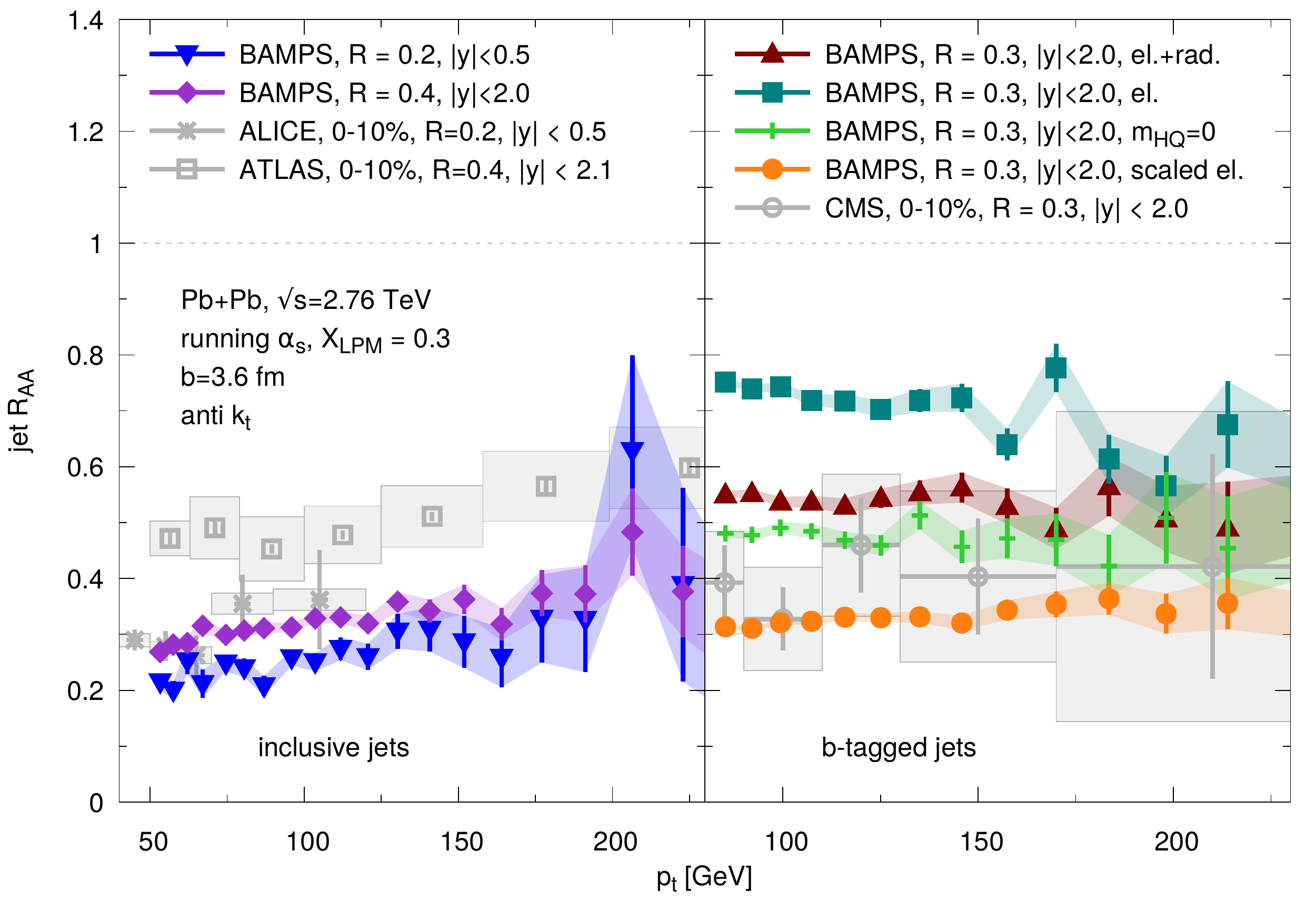}
\caption{Nuclear modification factor $R_{AA}$ of inclusive (left) and b-tagged reconstructed jets (right) calculated by \Bamps{} together with data of 0-10\% \PbPb{} collisions at $\sqrt{s}_{\rm LHC} = \SI{2.76}{\ATeV}$ \cite{ATLASCollaboration2014,ALICECollaboration2015b,CMSCollaboration2013}.}
\label{fig:jet_raa_hq_inclusive}
\end{figure}

Fig.~\ref{fig:jet_raa_hq_inclusive} shows the nuclear modification factor \Raa{} of inclusive (left panel) and b-tagged jets (right panel) calculated within \Bamps{} together with the corresponding experimental data at LHC. Obviously, the inclusive jet suppression resulting from gluon and light quark interactions within \Bamps{} is too strong in comparison with data for both employed radii $R=0.2$ and $R=0.4$. Since \Bamps{} simulations for the single hadron \Raa{} shows a realistic energy loss of the leading parton \cite{Uphoff2015a}, this finding shows that the radiated gluons are transported too far away from the respective jet axis within \Bamps{}. One possible reason for this strong jet suppression could be the current implementation of the LPM effect. The theta function in the radiative matrix elements effectively rejects collinear gluon emissions what potentially leads to broader angular distributions.

In the right panel of Fig.~\ref{fig:jet_raa_hq_inclusive} we present our results for the b-tagged jet suppression: While only elastic heavy quark processes are insufficient for a realistic jet suppression, the additional medium-induced gluon radiation off the b-quark in scenario "el.+rad." provides an important contribution to the medium transport out of the jet cones. Surprisingly, also the scaled elastic heavy quark interactions, lacking any additional medium-induced gluon radiation, even further enhance the b-tagged jet suppression. Both the modified screening via the parameter $\kappa$ and the $K$-factor lead to an increased elastic energy loss and thereby increase the probability for deflecting the b-quark to larger angles what results in smaller jet momenta. Both effects compensate the missing medium-induced gluon radiation off the bottom quark.

Furthermore, the b-tagged jet suppression of scenarios "el.+rad." and "el.+rad.,${\rm m}_{\rm HQ} = 0$" show a similar dependence, with a slightly stronger suppression for scenario "el.+rad.,${\rm m}_{\rm HQ} = 0$". This finding suggests that the different suppression of light and heavy flavor jets is mainly caused by the different initial shower evolution within \Pythia{}. As demonstrated in Ref.~\cite{Uphoff2014} and also in Refs.~\cite{Huang2013,Djordjevic2016} the finite mass of the bottom quark during the in-medium evolution plays at these high momenta only a minor role since coherence effects dominate the momentum loss.

\begin{figure}[tb]
\includegraphics[width=\columnwidth]{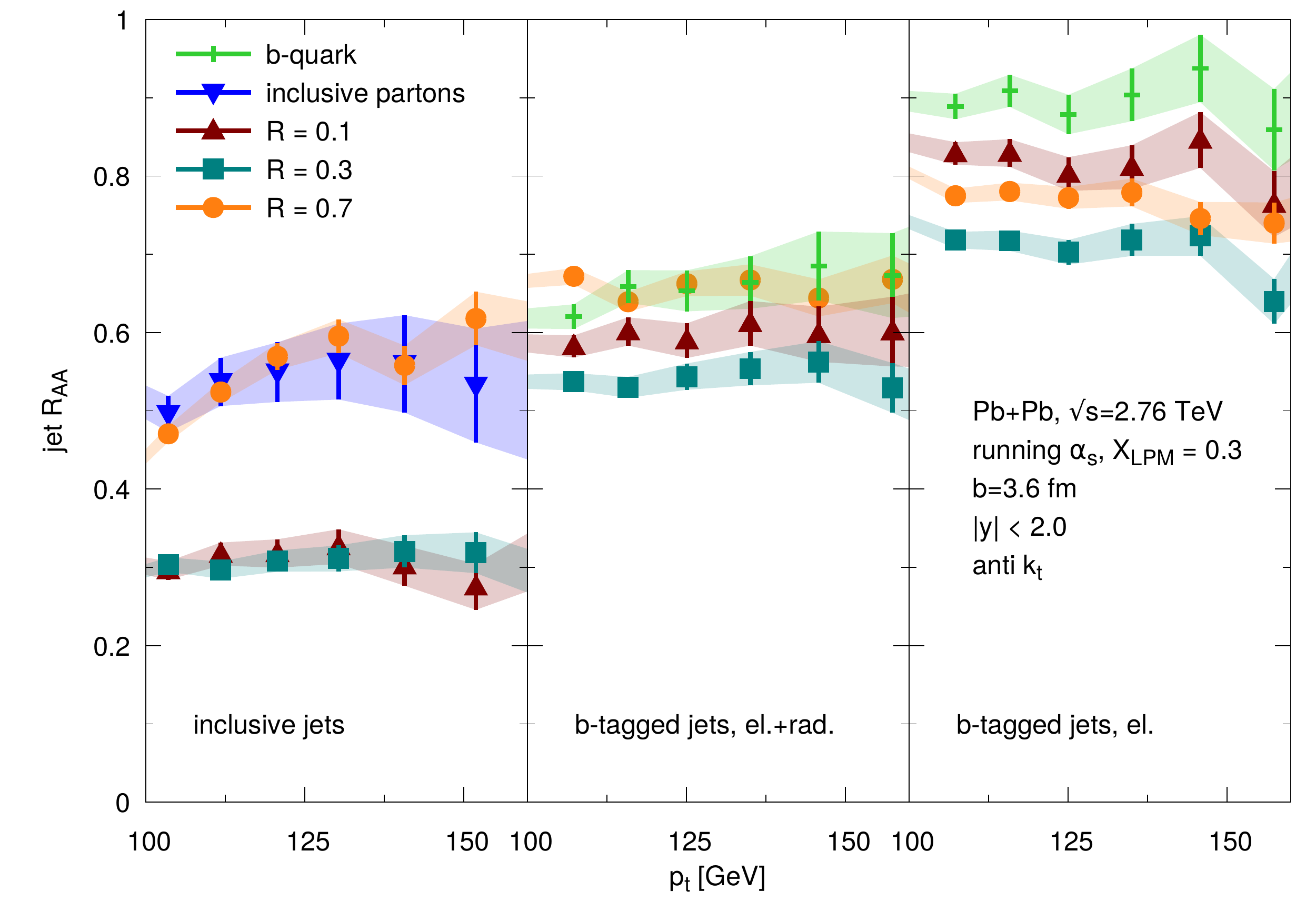}
\caption{$R$ dependence of the nuclear modification factor $R_{AA}$ of inclusive (left) and b-tagged jets (middle,right) for two different heavy quark energy loss scenarios.}
\label{fig:jet_raa_Rdependence}
\end{figure}

To further study the angular dependence of momentum transport out of the jet cones, we show in Fig.~\ref{fig:jet_raa_Rdependence} the nuclear modification factor \Raa{} of both inclusive (left panel) and b-tagged jets (middle and right panel) for different jet radii $R \in [0.1;0.3;0.7]$ in the range $p_t=\SIrange{100}{170}{\giga\electronvolt}$. The inclusive jet spectra show at first an expectable behavior: A larger radius $R=0.7$ leads to a decreased jet suppression since gluons transported away from the jet axis are recollected within the jets. Consequently, the lack of these gluons closer to the jet axis leads to a stronger suppression for the smaller radii $R=0.1$ and $R=0.3$. Furthermore, the jet \Raa{} within \Bamps{} is similar for both smaller radii demonstrating again that the medium-modification of inclusive jets is dominated by the large $R$ region. 
On the contrary, when comparing the jet suppression to the $R_{AA}$ of single inclusive partons (corresponding to the limit $R=\num{0}$), the suppression of spectra decreases again comparable to the jet $R_{AA}$ with $R=0.7$ since the momentum loss of a \enquote{bare} quark/gluon is smaller than of a reconstructed jet where also missing shower gluons contribute to the jet momentum loss. 

As presented in the middle and right panel of Fig.~\ref{fig:jet_raa_Rdependence} the R dependence of b-tagged jets shows a slightly different behavior: While the suppression at $R=0.7$ is moderated again by recollecting gluons at large angles, we predict already for the small radius $R=0.1$ a weaker jet suppression of b-tagged jets relative to the larger radius $R=0.3$ for both shown scenarios. This trend is further confirmed by comparing with the bare b-quark suppression (green line). The difference between the smaller radius $R=0.1$ and the larger radius $R=0.3$ is even more pronounced in the scenario \enquote{el.} where medium-induced radiation off the b-quark is neglected. Reason for this behavior is the different initial shower distribution of light and heavy flavor partons. In \pp{} collisions both light and heavy partons from the hard interactions radiate gluons due to their large virtuality. However, since the mass of the bottom quark suppresses gluon emissions at small angles, the initial shower of a bottom quark differs from a light parton shower since its core mainly consists of the leading heavy quark. Therefore when reconstructing jets with a high resolution (small radius) in \pp{} and subsequently in heavy-ion collisions, one measures almost only the modification of the leading bottom quark and not its associated parton shower. In contrast, the core of a jet initiated by a light parton is occupied (light quark initiated) or even dominated (gluon initiated) by shower gluons. Therefore the subsequent transport of these gluons out of the jet cones additionally contributes to the jet momentum loss and thereby to the jet suppression of inclusive jets already at small $R$. This leads to the mentioned difference in the suppression of inclusive and b-tagged jets reconstructed at small $R$.

Furthermore, when decreasing the resolution (increasing the radius to e.g. $R=\num{0.3}$) also the b-tagged jets become sensitive to the transport of gluons at intermediate angles out of the jet cones what results in a stronger suppression compared to smaller radii (e.g. $R=0.1$). This is even more pronounced in scenario \enquote{el.} where no additional medium-induced gluon radiation off the b-quark is considered and therefore a larger difference between $R=0.1$ and $R=0.3$ is visible. On the contrary, jets initiated by a light parton lack this effect since the gluons of the associated shower are transported to angles larger than $R=0.3$ resulting in a similar jet suppression from $R=0.1$ to $R=0.3$. Only at a small resolution (large radius) the shower gluons of both light and heavy flavor initiated showers are recollected to the jets so that the suppression of both inclusive and b-tagged jets decreases. Moreover, when comparing the suppression of b-tagged jets at $R=0.7$ for both scenarios one also can infer that the radiative processes lead to larger differences in the region between $R=0.3$ and $R=0.7$ than the only elastic bottom quark interactions, where the suppression of b-tagged jets with $R=0.7$ is closer to the suppression of b-tagged jets with $R=0.3$. This again shows that the radiation off the bottom quark additionally modifies the underlying jet structure.

\begin{figure}[tb]
\includegraphics[width=\columnwidth]{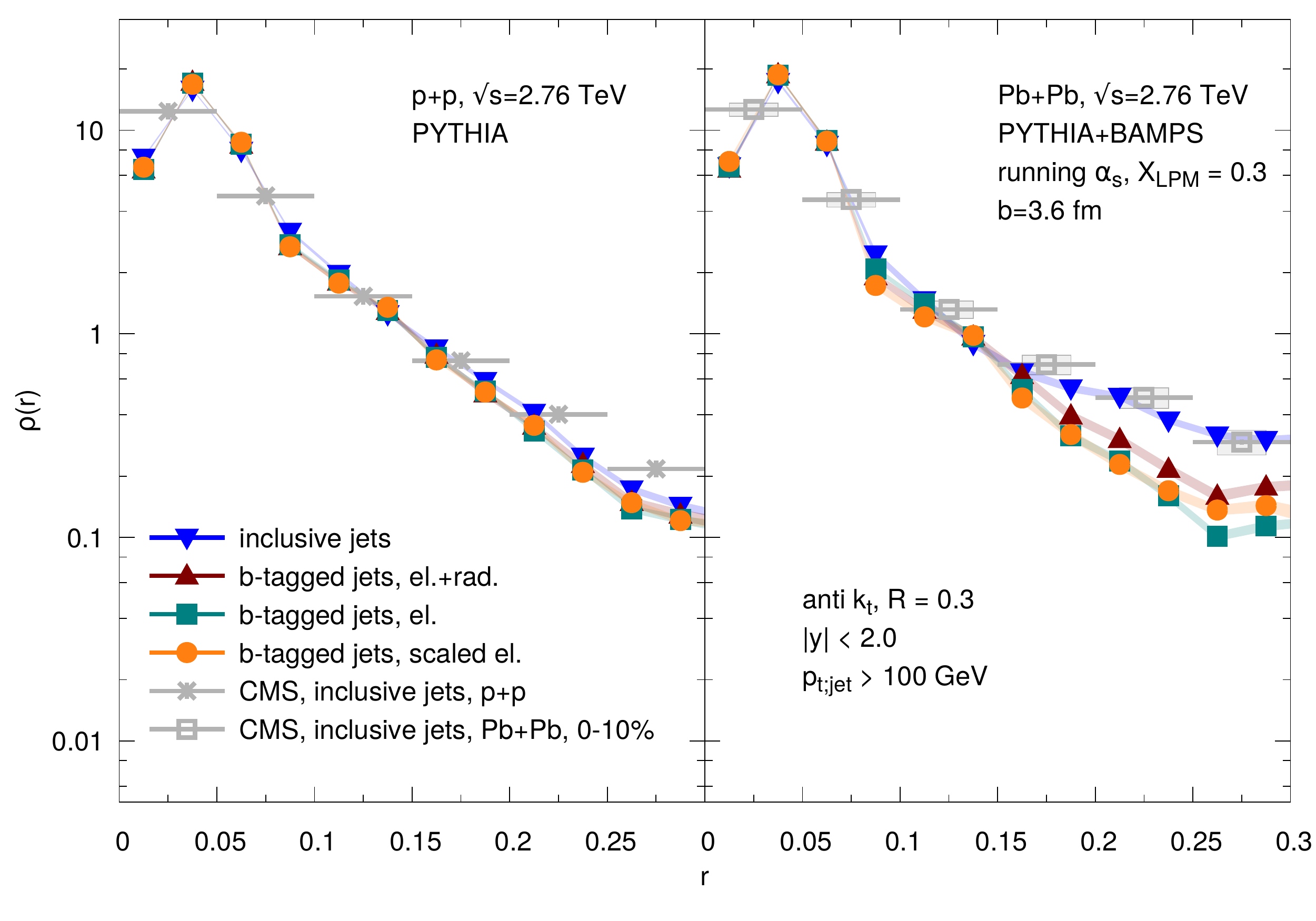}
\caption{Jet shapes of inclusive and b-tagged jets with $p_{\rm t;jet} > \SI{100}{\giga\electronvolt}$ calculated by \Bamps{} with $\delta r = \num{0.025}$ together with data of \pp{} and 0-10\% \PbPb{} collisions both at $\sqrt{s}_{\rm LHC} = \SI{2.76}{\ATeV}$.}
\label{fig:jet_shapes}
\end{figure}

In the following, we further develop our picture of jet quenching by studying the medium-modification of jet shapes $\rho(r)$ (cf. eq.~\ref{eq:jetshapes}). In contrast to the results for \Raa{}, jet shapes are more sensitive to the resolution with which the jet axis can be determined since single particles are correlated with the reconstructed jets. Therefore we reconstruct jets from discretized cells in $\Delta \phi$-$\Delta y$ with bin width \num{0.1} instead from single partons in order to mimic the finite resolution effects of experimental calorimeters \cite{CMSCollaboration2013,Senzel2015a}. By comparing with jet shapes based on single partons (not shown) this finite resolution effect leads to \enquote{wobbling} jet axes that result in broader jet shape distributions around $r=0$.

Fig.~\ref{fig:jet_shapes} shows the normalized jet shape distributions from \Pythia{} before (left panel) and after evolving within \Bamps{} (right panel) for inclusive jets and the different b-tagged jet scenarios in comparison with data for \pp{} and 0-10\% central \PbPb{} collisions both at $\sqrt{s}_{\rm LHC} = \SI{2.76}{\ATeV}$. In \pp{} collisions the distributions of inclusive and b-tagged jets are comparable with a slightly steeper distribution for the b-tagged jets. Reasons for this are the mentioned smearing of the jet axis by calorimeter effects and the similar distribution of inclusive jets and b-tagged jets emerging from gluon splitting processes. After traversing the \Bamps{} medium both jet shape distributions become flatter at $r>0.1$ what is consistent with the experimental data for the inclusive jets. However, in contrast to the similar jet \Raa{} of inclusive and b-tagged jets, the jet shape distribution shows a significantly less fraction of jet momentum at $r>0.15$ for b-tagged than for inclusive jets. This is caused by the missing gluon radiation of a bottom quark within \Pythia{} which results in fewer gluons that can be transported to these angles by the medium. The medium-induced radiation of the bottom quark is insufficient for refilling this region as can be seen by comparing the jet shapes of the different b-tagged jet scenarios.

\begin{figure}[tb]
\includegraphics[width=\columnwidth]{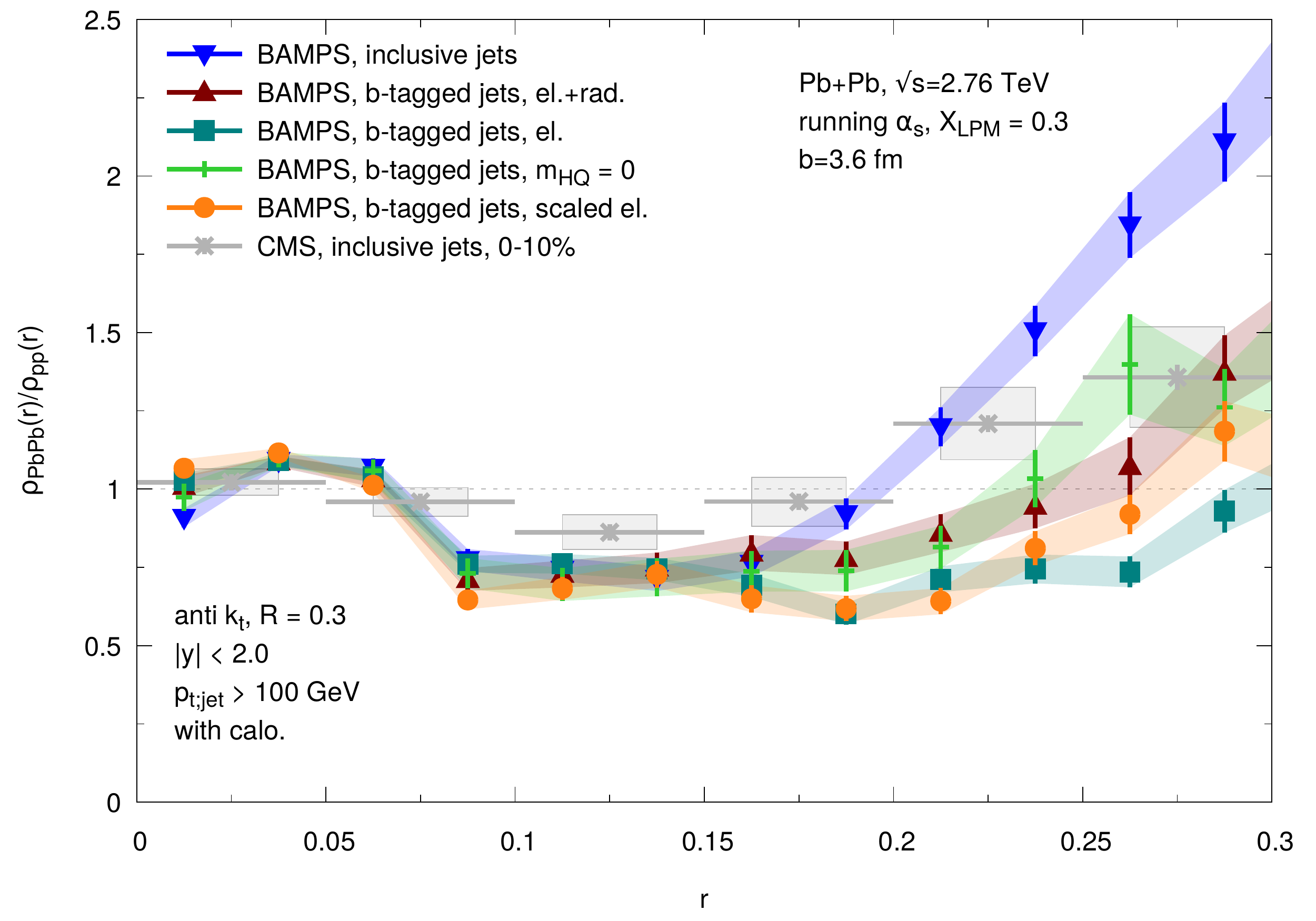}
\caption{Medium modification of jet shapes of inclusive and b-tagged jets with $p_{\rm t;jet} > \SI{100}{\giga\electronvolt}$ calculated by \Bamps{} with $\delta r = \num{0.025}$ together with data of 0-10\% \PbPb{} collisions at $\sqrt{s}_{\rm LHC} = \SI{2.76}{\ATeV}$.}
\label{fig:jet_shapes_ratio}
\end{figure}

This missing replenishment of gluons at intermediate angles $r>0.1$ can be further demonstrated in Fig.~\ref{fig:jet_shapes_ratio} where we show the ratio of jet shape distributions before and after the \Bamps{} evolution in comparison with data and thereby predict for the first time the medium-modification of b-tagged jet shapes. On one hand both the inclusive and b-tagged jet shapes show  the same behavior: an unmodified inner jet core and for $r\approx\SIrange{0.075}{0.15}{}$ an approx. 25\% depletion of momentum compared to \pp{} collisions. On the other hand, one finds a too strong enhancement for the distributions of inclusive jets at large angles $r>\num{0.25}$ that is consistent with the too strong suppression of inclusive jets within \Bamps{}. In contrast, all heavy quark scenarios show for $r > 0.15$ a significantly smaller modification. This is caused by a depletion resulting from shower gluons transported to larger angles that are not sufficiently refilled by medium-induced gluon radiation. Moreover, the different heavy quark interactions lead to a hierarchy of modification at larger $r>\num{0.25}$: The radiative heavy quark processes of scenario \enquote{el.+rad.} lead to the strongest enhancement of momentum at large $r$, while the elastic processes are less effective in transporting momentum to these large angles. Furthermore, the similarity of the scenarios \enquote{el.+rad.} and \enquote{el.+rad.,$m_{\rm HQ}=0$} suggest again that the jet shapes are mainly caused by the initial characteristic distribution around the b-quark and not by the finite mass within the \Bamps{} evolution. With increasing experimental statistics the differences between the elastic and radiative scenarios will hopefully be measurable and thereby provide evidence for the respective underlying energy loss. 

In summary, we compared results on the nuclear modification factor of reconstructed inclusive and b-tagged jets in ultra-relativistic heavy-ion collisions obtained from microscopic, non-equilibrium transport calculations. While the suppression of inclusive reconstructed jets seems to be strong in comparison with data from the LHC, the so far measured b-tagged jet \Raa{} does not allow a reliable discrimination between scaled elastic and elastic+radiative processes within \Bamps{}. Furthermore, we predicted the $R$ dependence of the b-tagged jet \Raa{} caused by the suppressed gluon radiation of b-quarks in vacuum and thereby a missing replenishment of gluons at intermediate angles relative to the jet axis. This finding was confirmed by a closer look at the medium modification of inclusive and b-tagged jet shapes. By predicting b-tagged jet shapes we find that their measurement should allow a reliable discrimination between different heavy quark energy loss mechanisms and thereby facilitates a further understanding of the mass dependence of jet quenching. Finally, we would like to emphasize that the presented results mainly follow from generic considerations and therefore should also, at least qualitatively, be observable in other pQCD energy loss models. 

\textbf{Acknowledgments}:                                                                                                                                                                                                     
FS thanks R. Bertens for helpful discussions regarding the experimental feasibility. This work was supported by the Bundesministerium f\"ur Bildung und Forschung (BMBF), the MOST, the NSFC under Grants No. 2014CB845400, No. 11275103, No. 11335005, HGS-HIRe, and the Helmholtz International Center for FAIR within the framework of the LOEWE program launched by the State of Hesse. Numerical computations have been performed at the Center for Scientific Computing (CSC).

//\bibliography{library}

\end{document}